# Grounded Symbols in the Brain
## Computational Foundations for Perceptual Symbol System

**Leonid Perlovsky and Roman Ilin**

Leonid Perlovsky, Harvard University and AFRL; leonid@seas.harvard.edu.
Roman Ilin, AFRL; roman.ilin@hanscom.af.mil.

## Abstract

We describe a mathematical models of grounded symbols in the brain. It also serves as a computational foundations for Perceptual Symbol System (PSS). This development requires new mathematical methods of dynamic logic (DL), which have overcome limitations of classical artificial intelligence and connectionist approaches. The paper discusses these past limitations, relates them to combinatorial complexity (exponential explosion) of algorithms in the past, and further to the static nature of classical logic. The new mathematical theory, DL, is a process-logic. A salient property of this process is evolution of vague representations into crisp. The paper first applies it to one aspect of PSS: situation learning from object perceptions. Then we relate DL to the essential PSS mechanisms of concepts, simulators, grounding, productivity, binding, recursion, and to the mechanisms relating grounded and amodal symbols. We discuss DL as a general theory describing the process of cognition on multiple levels of abstraction. We also discuss the implications of this theory for interactions between cognition and language, mechanisms of language grounding, and possible role of language in grounding abstract cognition. The developed theory makes experimental predictions, and will impact future theoretical developments in cognitive science, including knowledge representation, and perception-cognition interaction. Experimental neuroimaging evidence for DL and PSS in brain imaging is discussed as well as future research directions.

## 1. Introduction. PSS, challenge of computational model

Perceptual symbol system (PSS) grounds cognition in perception (Barsalou 1999). "Grounded cognition… rejects the standard view that amodal symbols represent knowledge in semantic memory" (Barsalou 1999). PSS emphasized the roles of simulation in cognition. "Simulation is the reenactment of perceptual, motor, and introspective states acquired during experience with the world, body, and mind… when knowledge is needed to represent a category (e.g., chair), multimodal representations captured during experiences… are reactivated to simulate how the brain represented perception, action, and introspection associated with it." Simulation is an essential computational mechanism in the brain. The best known case of these simulation mechanisms is mental imagery (e.g., Kosslyn 1980; 1994). According to PSS cognition supports action. Simulation is a central mechanism of PSS, yet rarely, if ever, they recreate full experiences. Using the mechanism of simulators, which approximately correspond to concepts and types in amodal theories, PSS implements the standard symbolic functions of type-token binding, inference, productivity, recursion, and propositions. Using these mechanisms PSS retains the symbolic functionality. "Thus, PSS is a synthetic approach that integrates traditional theories with grounded theories." (Barsalou 1999; 2005; 2007).

According to Barsalou, during the Cognitive Revolution in the middle of the last century, cognitive scientists were inspired by new forms of representation "based on developments in logic, linguistics, statistics, and computer science." They adopted amodal representations, such as feature lists, semantic networks, and frames (Barsalou & Hale 1993). Little empirical evidence supports amodal symbolic mechanisms (Barsalou 1999). It seems that amodal symbols were adopted largely because they promised to provide "elegant and powerful formalisms for representing knowledge, because they captured important intuitions about the symbolic character of cognition, and because they could be implemented in artificial intelligence." As we discuss in the next section, these promises were unfulfilled due to fundamental mathematical difficulties.

There is a number of past and ongoing developments of computational implementations of PSS (Cangelosi et al. 2000; Cangelosi & Riga 2006) and references therein. Yet, computational models for PSS (Barsalou 1999; 2007) require new mathematical methods different from traditional artificial intelligence, pattern recognition, or connectionist methods. The reason is that the traditional methods encountered combinatorial complexity (CC), an irresolvable computational difficulty, when attempting to model complex systems. Cognitive modeling requires learning combinations of perceptual features and objects or events (Perlovsky 1994; 1997; 2001; 2002a,b; 2006b; 2007a).

The aim of this article is to develop a realistic and scalable mathematical model of grounded symbols and formalization of PSS based on a new computational technique of dynamic logic, DL (Perlovsky 2006a,b). Although the developed mathematical formalism is quite general, here we first concentrate on just one example of PSS mechanism: a mathematical description of models and simulators for forming and enacting representations of situations (higher level symbols) from perceptions of objects (lower level symbols), and then we discuss its general applicability. In addition to simulators, we consider concepts, grounding, binding, dynamic aspect of PSS (DIPSS), abstract concepts, the mechanism of amodal symbols within PSS, and the role of logic.

Section 2 considers past mathematical difficulties, relates it to classical logic, and introduces a new computational technique of dynamic logic (DL), which overcomes past computational limitations. Whereas classical logic is a static logic of statements (e.g., "if A then B"), DL describes a process capable of modeling the main components of PSS, including simulators. Section 3 illustrates the important properties of DL. Section 4 illustrates how DL models essential mechanisms of PSS considering an example of learning situations from objects (a difficult problem due to its inherent combinatorial complexity). Section 5 discusses DL as a general mechanism of interacting bottom-up and top-down signals, applicable to all levels of cognitive processing., We discuss the representation of abstract concepts, the role of language, and how it is relevant to modeling PSS. Section 6 continues this discussion concentrating specifically on DL modeling amodal vs. perceptual symbols. Section 7 discusses experimental evidence confirming DL predictions of the mind mechanism, and formulates further predictions that could be tested experimentally in the near future. Section 8 describes future theoretical research as well as proposed verifiable experimental predictions of DL.

## 2.  Overcoming past mathematical difficulties

According to modern neuroscience, object perception involves bottom-up signals from sensory organs and top-down signals from internal mind's representations (memories) of objects. During perception, the mind matches subsets of bottom-up signals corresponding to objects with representations of object in the mind (and top-down signals). This produces object recognition; it activates brain signals leading to mental and behavioral responses (Grossberg 1982; Kosslyn 1994; Bar et al 2006; Schacter & Addis 2007). This section briefly summarizes mathematical development in artificial intelligence, pattern recognition, and other computational methods used in cognitive science for modeling brain-mind processes. We discuss the fundamental difficulties preventing mathematical modeling of perception, cognition, PSS, and the role of DL in overcoming these difficulties.

### 2.1.    Computational complexity since the 1950s

Developing mathematical descriptions of the very first recognition step in this seemingly simple association-recognition-understanding process has not been easy, a number of difficulties have been encountered during the past fifty years. These difficulties were summarized under the notion of combinatorial complexity (CC) (Perlovsky 1998b). CC refers to multiple combinations of bottom-up and top-down signals, or more generally to combinations of various elements in a complex system; for example, recognition of a scene often requires concurrent recognition of its multiple elements that could be encountered in various combinations. CC is computationally prohibitive because the number of combinations is very large: for example, consider 100 elements (not too large a number); the number of combinations of 100 elements is $100^{100}$, exceeding the number of all elementary particle events in life of the Universe; no computer would ever be able to compute that many combinations. Although, the story might sound "old," we concentrate here on those aspects of mathematical modeling of the brain-mind, which remain current and affect thinking in computational modeling and in cognitive science of many scientists today.

The problem of CC was first identified in pattern recognition and classification research in the 1960s and was named "the curse of dimensionality" (Bellman 1961). It seemed that adaptive self-learning algorithms and neural networks could learn solutions to any problem 'on their own', if provided with a sufficient number of training examples. The following decades of developing adaptive statistical pattern recognition and neural network algorithms led to a conclusion that the required number of training examples often was combinatorially large. This remains true about recent generation of algorithms and neural networks, which are much more powerful than those in the 1950s and 60s. Training had to include not only every object in its multiple variations, angles, etc., but also combinations of objects. Thus, self-learning approaches encountered CC of learning requirements.

Rule systems were proposed in the 1970's to solve the problem of learning complexity (Minsky 1975; Winston 1984). Minsky suggested that learning was a premature step in artificial intelligence; Newton "learned" Newtonian laws, most of scientists read them in the books. Therefore Minsky has suggested, knowledge ought to be input in computers "ready made" for all situations and artificial intelligence would apply these known rules. Rules would capture the required knowledge and eliminate a need for learning. Chomsky's original ideas concerning mechanisms of language grammar related to deep structure (Chomsky 1972) were also based on logical rules. Rule systems work well when all aspects of the problem can be predetermined. However in the presence of variability, the number of rules grew; rules became contingent on other rules and combinations of rules had to be considered. The rule systems encountered CC of rules.

In the 1980s, model systems were proposed to combine advantages of learning and rules-models by using adaptive models (Nevatia & Binford 1977; Bonnisone et al. 1991; Perlovsky 1987; 1988; 1991; 1994). Existing knowledge was to be encapsulated in models and unknown aspects of concrete situations were to be described by adaptive parameters. Along similar lines went the principles and parameters idea of Chomsky (1981). Fitting models to data (top-down to bottom-up signals) required selecting data subsets corresponding to various models. The number of subsets, however, is combinatorially large. A general popular algorithm for fitting models to the data, multiple hypothesis testing (Singer, Sea, & Housewright 1974) is known to face CC of computations. Model-based approaches encountered computational CC (N and NP complete algorithms).

## 2.2.    Logic, CC, and amodal symbols.

Amodal symbols and perceptual symbols described by PSS differ not only in their representations in the brain, but also in their properties that are mathematically modeled in this paper. This mathematically fundamental difference and its relations to CC of matching bottom-up and top-down signals are the subjects of this section.

The fundamental reasons for CC are related  to the use of formal logic by algorithms and neural networks (Perlovsky 1996a; 2001; Marchal 2005). Logic serves as a foundation for many approaches to cognition and linguistics; it underlies most of computational algorithms. But its influence extends far beyond, affecting cognitive scientists, psychologists, and linguists, who do not use complex mathematical algorithms for modeling the mind. All of us operate under the influence of formal logic that is more than 2000 years old, making a more or less conscious assumption that the mechanisms of logic serve as  the basis of our cognition. As discussed in section 7, our minds are unconscious about its illogical foundations. We are mostly conscious about a small part of the mind mechanisms, which is  approximately logical. Our intuitions, therefore, are unconsciously affected by the bias toward logic. Even when the laboratory data drive our thinking away from logical mechanisms we are having difficulties overcoming the

logical bias (Grossberg 1988; Bar et al 2006; Perlovsky 1997, 2000, 2001, 2004, 2006b, 2007c, 2009c, 2010b,c).

The relationships between logic, cognition, and language have been a source of longstanding controversy. The widely accepted story is that Aristotle founded logic as a fundamental mind mechanism, and only during the recent decades science overcame this influence. I would like to emphasize the opposite side of this story. Aristotle assumed a close relationship between logic and language. He emphasized that logical statements should not be formulated too strictly and language inherently contains the necessary degree of precision. According to Aristotle, logic serves to communicate already made decisions (Perlovsky 2007). The mechanism of the mind relating language, cognition, and the world Aristotle described as forms. Today we call similar mechanisms mental representations, or concepts, or simulators in the mind. Aristotelian forms are similar to Plato's ideas with a marked distinction, forms are dynamic: their initial states, before learning, are different from their final states of concepts (1995 / IV BCE). Aristotle emphasized that initial states of forms, forms-as-potentialities, are not logical (i.e. vague), but their final forms, forms-as-actualities, attained in the result of learning, are logical. This fundamental idea was lost during millennia of philosophical arguments. As discussed below this Aristotelian process of dynamic forms corresponds to Barsalou idea of PSS simulators, and in this paper we describe the mathematical model, DL, for this process.

The founders of formal logic emphasized a contradiction between logic and language. In the 19[th] century George Boole and the great logicians following him, including Gottlob Frege, Georg Cantor, David Hilbert, and Bertrand Russell (see Davis 2000, and references therein) eliminated the uncertainty of language from mathematics, and founded formal mathematical logic, the foundation of the current classical logic. Hilbert developed an approach named formalism, which rejected intuition as a matter of scientific investigation and formally defined scientific objects in terms of axioms or rules. In 1900 he formulated famous Entscheidungsproblem: to define a set of logical rules sufficient to prove all past and future mathematical theorems. This was a part of "Hilbert's program," which entailed formalization of the entire human thinking and language. Formal logic ignored the dynamic nature of Aristotelian forms and rejected the uncertainty of language. Hilbert was sure that his logical theory described mechanisms of the mind. "The fundamental idea of my proof theory is none other than to describe the activity of our understanding, to make a protocol of the rules according to which our thinking actually proceeds." (see Hilbert 1928). However, Hilbert's vision of formalism explaining mysteries of the human mind came to an end in the 1930s, when Gödel (1932/1994) proved internal inconsistency of formal logic. This development called Gödel theory is considered among most fundamental mathematical results of the previous century. Logic, that was believed to be a sure way to derive truths, turned out to be basically flawed. This is a reason why theories of cognition and language based on formal logic are inherently flawed.

There is a close relation between logic and CC. It turned out that combinatorial complexity of algorithms is a finite-system manifestation of the Gödel's theory (Perlovsky 1996a). If Gödelian theory is applied to finite systems (all practically used or discussed systems, such as computers and brain-mind, are finite), CC is the result, instead of the fundamental inconsistency. Algorithms matching bottom-up and top-down signals based on formal logic have to evaluate every variation in signals and their combinations as separate logical statements. A large number of combinations of these variations cause CC.

This general statement manifests in various types of algorithms in different ways. Rule systems are logical in a straightforward way, and the number of rules grows combinatorially. Pattern recognition algorithms and neural networks are related to logic in learning procedures: every training sample is treated as a logical statement ("this is a chair") resulting in CC of learning. Multivalued logic and fuzzy logic were proposed to overcome limitations related to logic (Zadeh 1965; Kecman 2001). Yet the mathematics of multivalued logic is no different in principle from formal logic (Perlovsky 2006b). Fuzzy logic uses logic to set a degree of fuzziness. Correspondingly, it encounters a difficulty related to the degree of fuzziness: if too much fuzziness is specified, the solution does not achieve a needed accuracy, and if too little, it becomes similar to formal logic. If logic is used to find the appropriate fuzziness for every model at every processing step, then the result is CC. The mind has to make concrete decisions, for example one either enters a room or does not; this requires a computational procedure to move from a fuzzy state to a concrete one. But fuzzy logic does not have a formal procedure for this purpose; fuzzy systems treat this decision on an ad-hoc logical basis.

Is logic still possible after Gödel's proof of its incompleteness? The contemporary state of this field was reviewed in (Marchal 2005). It appears that logic after Gödel is much more complicated and much less logical than was assumed by founders of artificial intelligence. CC cannot be solved within logic. Penrose thought that Gödel's results entail incomputability of the mind processes and testify for a need for new physics "correct quantum gravitation," which would resolve difficulties in logic and physics (Penrose 1994). An opposite position in (Perlovsky 2001; 2006a;b;c) is that incomputability of logic does not entail incomputability of the mind. These publications add mathematical arguments to Aristotelian view that logic is not the basic mechanism of the mind.

To summarize, various manifestations of CC are all related to formal logic and Gödel theory. Rule systems rely on formal logic in a most direct way. Even mathematical approaches specifically designed to counter limitations of logic, such as fuzzy logic and the second wave of neural networks (developed after the 1980s) rely on logic at some algorithmic steps. Self-learning algorithms and neural networks rely on logic in their training or learning procedures: every training example is treated as a separate logical statement. Fuzzy logic systems rely on logic for setting degrees of fuzziness. CC of mathematical approaches to the mind is related to the fundamental inconsistency of logic. Therefore logical inspirations, leading early cognitive scientists to amodal brain mechanisms, could not realize their hopes for mathematical models of the brain-mind.

Why did the outstanding mathematicians of the 19[th] and early 20[th] c. believed in logic to be the foundation of the mind? Even more surprising is the belief in logic after Gödel. Gödelian theory was long recognized among most fundamental mathematical results of the 20[th] c. How is it possible that outstanding minds, including founders of artificial intelligence, and many cognitive scientists and philosophers of mind insisted that logic and amodal symbols implementing logic in the mind are adequate and sufficient? The answer, in our opinion, might be in the "conscious bias." As we discuss in section 7, non-logical operations making up more than 99.9% of the mind functioning are not accessible to consciousness (Bar et al 2006). However, our consciousness functions in a way that makes us unaware of this. In subjective consciousness we usually experience perception and cognition as logical. Our intuitions are "consciously biased." This is why amodal logical symbols, which describe a tiny fraction of the mind mechanisms, have seemed to many the foundation of the mind (Grossberg 1988; Bar et al 2006; Perlovsky 1997, 2000, 2001, 2002b, 2004, 2006a,b, 2007c, 2009c, 2010b,c).

Another aspect of logic relevant to PSS is that it lacks dynamics; it is about static statements such as "this is a chair." Classical logic is good at modeling structured statements and relations, yet it misses the dynamics of the mind and faces CC, when attempts to match bottom-up and top-down signals. The essentially dynamic nature of the mind is not represented in mathematical foundations of logic. Dynamic logic discussed in the next section is a logic-process. It overcomes CC by automatically choosing the appropriate degree of fuzziness-vagueness for every mind's concept at every moment. DL combines advantages of logical structure and connectionist dynamics. This dynamics mathematically represents the learning process of Aristotelian forms (which are different from classical logic as discussed in this section) and serves as a foundation for PSS concepts and simulators.

## 2.3.    Dynamic logic-process

DL models perception as an interaction between bottom-up and top-down signals (Perlovsky 2001; 2006a,b; 2007c; 2009c; 2010c). This section concentrates on the basic relationship between the brain processes and the mathematics of DL. To concentrate on this relationship, we much simplify the discussion of the brain structures. We discuss visual recognition of objects as if the retina and the visual cortex each consists of a single processing layer of neurons where recognition occurs (which is not true, detailed relationship of the DL process to brain is considered in given references). Perception consists of the association-matching of bottom-up and top-down signals. Sources of top-down signals are mental representations, memories of objects created by previous simulators (Barsalou 1999); these representations model the patterns of bottom-up signals. In this way they are concepts (of objects), symbols of a higher order than bottom-up signals; we call them concepts or models. In perception processes the models are modified by learning and new models are formed; since an object is never encountered exactly the same as previously, perception is always a learning process. The DL processes along with concept-representations are mathematical models of the PSS simulators. The bottom-up signals, in this simplified discussion, are a field of neuronal synapse activations in visual cortex. Sources of top-down signals are mental representation-concepts or, equivalently, model-simulators (for short, models; please notice this dual use of the word model, we use "models" for mental representation-simulators, which match-model patterns in bottom-up signals; and we use "models" for mathematical modeling of these mental processes). Each mental model-simulator projects a set of priming, top-down signals, representing the bottom-up signals expected from a particular object. Mathematical models of mental models-simulators characterize these mental models by parameters. Parameters describe object position, angles, lightings, etc. (In case of learning situations considered later, parameters characterize objects and relations making up a situation.) To summarize this highly simplified description of a visual system, the learning-perception process "matches" top-down and bottom-up activations by selecting "best" mental models-simulators and their parameters and fitting them to the corresponding sets of bottom-up signals. This DL process mathematically models multiple simulators running in parallel, each producing a set of priming signals for various expected objects.

Mathematical criteria of the "best" fit between bottom-up and top-down signals were given in (Perlovsky 2001; 2006a,b). They are similar to probabilistic or informatics measures. In the first case they represent probabilities that the given (observed) data or bottom-up signals corresponds to representations-models (top-down signals) of particular objects. In the second case they represent information contained in representations-models about the observed data (in other words, information in top-down signals about bottom-up signals). These similarities are maximized over the model parameters. Results can be interpreted correspondingly as a maximum likelihood that models-representations fit sensory signals, or as maximum information in models-representations about the bottom-up signals. Both similarity measures account for all expected models and for all combinations of signals and models. Correspondingly, a similarity contains a large number of items, a total of $M^N$, where M is a number of models and N is a number of signals; this huge number is the cause for the combinatorial complexity discussed previously.

Maximization of a similarity measure is a mathematical model of an unconditional drive to improve the correspondence between bottom-up and top-down signals (representations-models). In biology and psychology it was discussed as curiosity, cognitive dissonance, or a need for knowledge since the 1950s (Harlow 1950; Festinger 1957; Cacioppo & Petty 1982). This process involves knowledge-related emotions evaluating satisfaction of this drive for knowledge (Grossberg & Levine 1987; Perlovsky 2001, 2000c, 2006a,b; 2007c; Perlovsky, Bonniot-Cabanac, & Cabanac 2010). In computational intelligence it is even more ubiquitous, every mathematical learning procedure, algorithm, or neural network maximizes some similarity measure. In the process of learning, mental concept-models are constantly modified. From time to time a system forms a new concept, while retaining an old one as well; alternatively, old concepts are sometimes merged or discarded.

The DL learning process, let us repeat, consists in estimating parameters of concept-models (mental representations) and associating subsets of bottom-up signals with top-down signals originating from these models-concepts by maximizing a similarity. Although a similarity contains combinatorially many items, DL maximizes it without combinatorial complexity (Perlovsky 1996b; 1997; 2001; 2006a,b; 2007c; 2010c) as follows. First, vague-fuzzy association variables are defined, which give a measure of correspondence between each signal and each model. They are defined similarly to the a posteriori Bayes probabilities, they range between 0 and 1, and as a result of learning they converge to the probabilities, under certain conditions. Often the association variables are close to bell-shapes.

The DL process is defined by a set of differential equations given in the above references; together with models discussed later it gives a mathematical description of the PSS simulators. To keep the paper self-consistent we summarize these equations in Appendix 1. Those interested in mathematical details can read the Appendix. However, basic principles of DL can be adequately understood from a conceptual description and examples in this and following sections. As a mathematical model of perception-cognitive processes, DL is a process described by differential equations given in the Appendix; in particular, fuzzy association variables f associate bottom-up signals and top-down models-representations. Among unique DL properties is an autonomous dependence of association variables on models-representations: in the processes of perception and cognition, as models improve and become similar to patterns in the bottom-up signals, the association variables become more selective, more similar to delta-functions. Whereas initial association variables are vague and associate near all bottom-up signals with virtually any top-down model-representations, in the processes of perception and cognition association variables are becoming specific, "crisp", and associate only appropriate signals. This we call a process "from vague to crisp." (The exact mathematical definition of crisp corresponds to values of f = 0 or 1; values of f in between 0 and 1 correspond to various degrees of vagueness.)

DL processes mathematically model PSS simulators and not static amodal signals. Another unique aspect of DL is that it explains how logic appears in the human mind; how illogical dynamic PSS simulators give rise of classical logic, and what is the role of amodal symbols. This is discussed throughout the paper, and also in specific details in section 6.

An essential aspect of DL, mentioned above, is that associations between models and data (top-down and bottom-up signals) are uncertain and dynamic; their uncertainty matches uncertainty of parameters of the models and both change in time during perception and cognition processes. As the model parameters improve, the associations become crisp. In this way the DL model of simulator-processes avoids combinatorial complexity because there is no need to consider separately various combinations of bottom-up and top-down signals. Instead, all combinations are accounted for in the DL simulator-processes. Let us repeat, that initially, the models do not match the data. The association variables are not the narrow logical variables 0, or 1, or nearly logical, instead they are wide functions (across top-down and bottom-up signals). In other words, they are vague, initially they take near homogeneous values across the data (across bottom-up and top-down signals); they associate all the representation-models (through simulator processes) with all the input signals (Perlovsky 2001; 2006b; 2009c). Here we conceptually describe the DL process as applicable to visual perception, taking approximately 160 ms, according to the reference below. Gradually, the DL simulator-processes improve matching, models better fit data, the errors become smaller, the bell-shapes concentrate around relevant patterns in the data (objects), and the association variables tend to 1 for correctly matched signal patterns and models, and 0 for others. These 0 or 1 associations are logical decisions. In this way classical logic appears from vague states and illogical processes. Thus certain representations get associated with certain subsets of signals (objects are recognized and concepts formed logically or approximately logically). This process "from vague-to-crisp" that matches bottom-up and top-down signals has been independently conceived and demonstrated in brain imaging research to take place in human visual system (Bar, Kassam, Ghuman, Boshyan, Schmid, Dale, Hamalainen, Marinkovic, Schacter, Rosen, & Halgren 2006). Thus DL models PSS simulators, describes how logic appears from illogical processes, and actually models perception mechanisms of the brain-mind.

Mathematical convergence of the DL process was proven in (Perlovsky 2001). It follows that the simulator-process of perception or cognition assembles objects or concepts among bottom-up signals, which are most similar in terms of the similarity measure. Despite a combinatorially large number of items in the similarity, a computational complexity of DL is relatively low, it is linear in the number of signals, and therefore could be implemented by a physical system, like a computer or brain.

## 2.4.    Example of DL, object perception in noise

The purpose of this section is to illustrate the DL processes, multiple simulators running in parallel as described above; our purpose here is not to illustrate DL-PSS relations. Therefore we use a simple example, still unsolvable by other methods, (mathematical details are omitted, they could be found in Linnehan, Mutz, Perlovsky, Weijers, Schindler, & Brockett 2003). In this example, DL searches for patterns in noise. Finding patterns below noise can be an exceedingly complex problem. If an exact pattern shape is not known and depends on unknown parameters, these parameters should be found by fitting the pattern model to the data. However, when the locations and orientations of patterns are not known, it is not clear which subset of the data points should be selected for fitting. A standard approach for solving this kind of problem, which has already been mentioned, is multiple hypotheses testing (Singer, Sea, & Housewright 1974); this algorithm exhaustively searches all logical combinations of subsets and models and faces combinatorial complexity. In the current example, we are looking for 'smile' and 'frown' patterns in noise shown in Fig.1a without noise, and in Fig.1b with noise, as actually measured (object signals are about 2-3 times below noise and cannot be seen).

To apply DL to this problem, we used DL equations given in the Appendix. Specifics of this example are contained in models. Several types of models are used: parabolic models describing 'smiles' and 'frown' patterns (unknown size, position, curvature, signal strength, and number of models), circular-blob models describing approximate patterns (unknown size, position, signal strength, and number of models), and noise model (unknown strength). Exact mathematical description of these models is given in the reference cited above.

The image size in this example is 100x100 points (N = 10,000 bottom-up signals, corresponding to the number of receptors in an eye retina), and the true number of models is 4 (3+noise), which is not known. Therefore, at least M = 5 models should be fit to the data, to decide that 4 fits best. This yields complexity of logical combinatorial search, $M^N = 10^{5000}$; this combinatorially large number is much larger than the size of the Universe and the problem was considered unsolvable. Fig. 1 illustrates DL operations: (a) true 'smile' and 'frown' patterns without noise; (b) actual image available for recognition; (c) through (h) illustrates the DL process, they show improved models at various steps of solving DL eqs.(A3), total of 22 steps (noise model is not shown; figures (c) through (h) show association variables, f, for blob and parabolic models). By comparing (h) to (a) one can see that the final states of the models match patterns in the signal. Of course, DL does not guarantee finding any pattern in noise of any strength. For example, if the amount and strength of noise would increase ten-fold, most likely the patterns would not be found (this would provide an example of "falsifiability" of DL; however more accurate mathematical description of potential failures of DL algorithms is considered later). DL reduced the required number of computations from combinatorial $10^{5000}$ to about $10^9$. By solving the CC problem DL was able to find patterns under the strong noise. In terms of signal-to-noise ratio this example gives 10,000% improvement over the previous state-of-the-art. (In this example DL actually works better than human visual system; the reason is that human brain is not optimized for recognizing these types of patterns in noise).

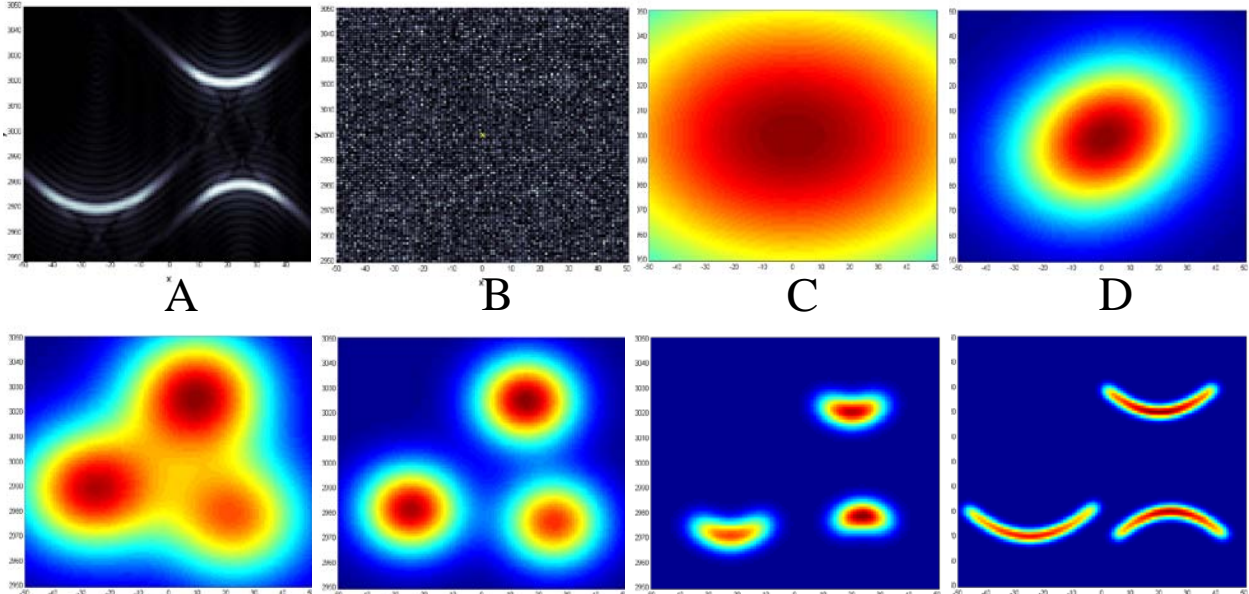

<div style="text-align: center">A      B      C      D</div>

<div style="text-align: center">E      F      G      H</div>

uncertainty of knowledge; (d) through (h) show improved models at various steps of DL (eq.(3) are solved in 22 steps). Between stages (d) and (e) the algorithm tried to fit the data with more than one model and decided, that it needs three blob-models to 'understand' the content of the data. There are several types of models: one uniform model describing noise (it is not shown) and a variable number of blob-models and parabolic models, which number, location, and curvature are estimated from the data. Until about stage (g) the algorithm 'thought' in terms of simple blob models, at (g) and beyond, the algorithm decided that it needs more complex parabolic models to describe the data. Iterations stopped at (h), when similarity (1) stopped increasing.

The main point of this example is that DL simulator-perception is a process "from vague-to-crisp," similar to visual system processes demonstrated in (Bar et al. 2006; in that publication **authors refer as "low spatial frequency" to what we call "vague" in Fig.1).**

We would also like to take this moment to continue the arguments from sections 2.1, 2.2, and to emphasize that DL is a fundamental and revolutionary improvement in mathematics (Perlovsky 2009a; 2010c); it was recognized as such in mathematical and engineering communities; it is the only mathematical theory that have suggested vague initial states; it has been developed for over 20 years; yet it might not be well known in cognitive science community. Those interested in a large number of mathematical and engineering applications of DL could consult given references and references therein. Here we would like to address two specific related concerns, first, if the DL algorithms are falsifiable, second, a possibility that Fig. 1 example could be "lucky" or "erroneous." We appreciate that some readers could be skeptical about 10,000% improvement over the state of the art. In mathematics there is a standard procedure for establishing average performance of detection (perception) and similar algorithms. It is called 'operating curves' and it takes not one example, but tens of thousands examples, randomly varying in parameters, initial conditions, etc. The results are expressed in terms of probabilities of correct and incorrect algorithm performance (this is an exact mathematical formulation of the idea of "falsifiability" of an algorithm). These careful procedures demonstrated that Fig. 1 represents an average performance of the DL algorithm (Perlovsky 2001).

## 3. DL of PSS: perceptual cognition and simulators

Section 2.4 illustrated DL for recognition of simple objects in noise, a case complex and unsolvable for prior state-of-the-art algorithms, still too simple to be directly relevant for PSS. Here we consider a problem of situation learning, assuming that object recognition has been solved. In computational image recognition this is called "situational awareness" and it is a long-standing unsolved problem. The principled difficulty is that every situation includes many objects that are not essential for recognition of this specific situation; in fact there are many more "irrelevant" or "clutter" objects than relevant ones. Let us dwell on this for a bit. Objects are spatially-limited material things perceptible by senses. A situation is a collection of contextually related objects that tend to appear together and are perceived as meaningful, e.g., an office, a dining room. The requirement for contextual relations and meanings makes the problem mathematically difficult. Learning contexts comes along with learning situations; it reminds of the problem of a chicken and egg. We subliminally perceive many objects, most of which are irrelevant, e.g. a tiny scratch on a wall, which we learn to ignore. Combinations of even a limited number of objects exceed what is possible to learn in a single lifetime as meaningful situations and contexts (e.g. books on a shelf) from random sets of irrelevant objects (e.g. a scratch on a wall, a book, and a pattern of tree branches in a window). Presence of hundreds (or even dozens) irrelevant objects makes learning by a child of mundane situations a mathematical mystery. In addition, we constantly perceive large numbers of different objects and their combinations, which do not correspond to anything worth learning and we successfully learn to ignore them.

An essential part of learning-cognition is to learn which sets of objects are important for which situations (contexts). The key mathematical property of DL that made this solution possible, same as in the previous section, is a process "from vague-to-crisp." Concrete crisp models-representations of situations are formed from vague models in the process of learning (or cognition-perception). We illustrate below how complex symbols, situations, are formed by situation-simulators from simpler perceptions, objects, which are simpler perceptual symbols, being formed by simulators at "lower" levels of the mind, comparative to "higher" situation-simulators. Situation-simulators operate on PSS representations of situations, which are dynamic and vague assemblages of situations from imagery (and other modalities), bits and pieces along with some relations among them perceived at lower levels. These pieces and relations may come from different past perceptions, not necessarily from a single perceptual mode, and not necessarily stored in a contiguous parts of the brain. The dynamic process of DL-PSS-simulation, which assembles these bits into situations attempting to match those before the eyes, is mostly unconscious. We will discuss in details in section 6 that these are perceptual symbols as described in (Barsalou 1999). DL mathematically models PSS simulators (Barsalou 1999), processes that match bottom-up perceptions with top-down signals, assemble symbols in cognition-perception, and assemble conceptual representations by recreating patterns of activation in sensorimotor brain areas (as discussed later in the paper). An essential mechanism of DL cognition-perception is a process of simulation of perceptual imagination-cognitions; these situation-symbols are simulated from simpler perceptions-objects (we repeat that these simulations-imaginations are not limited to imagery, and are mostly unconscious). And the same mechanism can simulate plans and more complex abstract thoughts, as discussed in later sections. Thus, in the following sections we demonstrate that DL mathematically models PSS simulators, in this case simulators of situations.

## 3.1.    DL formulation

In a simplified problem considered here, the task is for an intelligent agent (a child) to learn to recognize certain situations in the environment; while it is assumed that a child has learned to recognize objects. In real life a child learns to recognize situations, to some extent, in parallel with recognizing objects. But for simplicity of the illustration examples and discussions below, we consider a simplified case of objects being already known. For example, situation "office" is characterized by the presence of a chair, a desk, a computer, a book, a book shelf. Situation "playground" is characterized by the presence of a slide, a sandbox, etc. The principal difficulty is that many irrelevant objects are present in every situation. (This child learning is no different mathematically from an adult recognition.)

In the example below, $D_o$ is the total number of objects that the child can recognize in the world (it is a large number). In every situation he or she perceives $D_p$ objects. This is a much smaller number compared to $D_o$. Each situation is also characterized by the presence of $D_s$ objects essential for this situation ($D_s < D_p$). Normally nonessential objects are present and $D_s$ is therefore less than $D_p$. The sets of essential objects for different situations may overlap, with some objects being essential to more than one situation. We assume that each object is encountered in the scene only once. This is a minor and nonessential simplification, e.g. we may consider a set of similar objects as a new object. For example, "book" is an object and "books" is another object referring to more than one book.

The real life learning is sequential as a child is exposed to situations one at a time. Again, DL can handle this, but in this paper we consider the data about all the situations available at the time of learning.

Following (Ilin & Perlovsky 2009) a situation can be mathematically represented as a vector in the space of all objects, $\mathbf{X}_n = (x_{n1}, \ldots x_{ni}, \ldots x_{nDo})$. If the value of $x_{ni}$ is one the object i is present in the situation n and if $x_{ni}$ is zero, the corresponding object is not present. Since $D_o$ is a large number, $\mathbf{X}_n$ is a large binary vector with most of its elements equal to zero. A situation model is characterized by parameters, a vector of probabilities, $\mathbf{p}_m = (p_{m1}, \ldots p_{mi}, \ldots p_{mDo})$. Here $p_{mi}$ is the probability of object i being part of the situation m. Thus a situation model contains $D_o$ unknown parameters. Estimating these parameters constitutes learning. We would like to emphasize that although notations like $x_{ni}$, $p_{mi}$ might look like amodal symbols, such an interpretation would be erroneous. Correct interpretation of notations in a mathematical model depends on what actual physical entities are referred to by the notations. These notations refer to neural signals, elements from which simulator-processes assemble symbols of a higher level. As discussed, for simplicity of presentation of the results, we assume that lower-level simulator-processes responsible for object recognition have already run their course, and objects have been recognized at a lower level, therefore $x_{ni}$ are 0 or 1. Given mathematical formulation can use dynamic signals $x_{ni}$, parts of object-recognition simulators. We remind that simulators of interest in this example are situations; in addition to $x_{ni}$ these simulators involve dynamic neural signals referred by $p_{mi}$.

These are constituent signals of the ongoing simulator processes at the considered level of situations, which learn to recognize situations, symbols at a higher level (relative to objects). (We continue this discussion below in this section and also in section 6).

We model the elements of vector $\mathbf{p}_m$ as independent (this is not essential for learning, if presence of various objects in a situation actually is correlated, this would simplify learning, e.g. perfect correlation would make it trivial). Correspondingly, conditional probability of observing vector $\mathbf{X}_n$ in a situation m is then given by the standard formula (Jaynes 2003).

$$l(\mathbf{X}(n) \mid \mathbf{M}_m(n)) = \prod_{i=1}^{Do} p_{mi}^{X_{ni}}(1 - p_{mi})^{(1-X_{ni})} \tag{1}$$

Consider N perceptions a child was exposed to (N includes real "situations" and "irrelevant" random ones); among them most perceptions were "irrelevant" corresponding to observing random sets of objects, and M-1 "real" situations, in which $D_s$ objects were repeatedly present. All random observations we model by 1 model ("noise"); assuming that every object has an equal chance of being randomly observed in noise (which again is not essential) the probabilities for this noise model, m=1, are $p_{1i}$=0.5 for all i. Thus we define M possible sources for each of the N observed situations.

The total likelihood-similarity for our M models (M-1 "real" and 1 noise) is given by the same equation as similarity in the previous example (Perlovsky 2006b, also Appendix 1). And the same DL equations maximize it over the parameters, which in this case are the probabilities of objects constituting various situations. Specifics of this case of situations are given by models, which exact form is given by eq.(1). The general DL equations given in the Appendix can be significantly simplified in this case (Ilin & Perlovsky 2009) and we will describe them now in details. The DL is an iterative process, it involves a sequence of interactions between bottom-up and top-down signals; this sequence involves first, association variables connecting these signals, and second, parameter update equations, which improve parameter values in this interaction-learning. We use shorthand notations l(n|m) for conditional probabilities (1). The association variables connecting bottom-up signals n with top-down projections-simulators m are given by the general equation, as in Appendinx 1,

$$f(m|n) = r(m) \, l(n|m) \,/ \sum_{m' \in M} r(m') \, l(n|m'). \tag{2}$$

For intuitive understanding we point out that these association variables are different from eq.(1) in that they are normalized by the denominator, the sum of l(m|n) for a given bottom-up signal over all active simulators m. Whereas l(m|n) could vary greatly in their values, f(m|n) vary between 0 and 1. Also, eq.(2) contain parameters r(m), which are needed for the following reason: it is convenient to define conditional probabilities (1) assuming the simulator m actually is present and active; therefore r(m) are needed to define the actual probability of the simulator m in the current process. These association variables (2) are defined similarly to the a posteriori Bayesian probabilities that a bottom-up signal n comes from a situation-simulator m. Yet they are not probabilities as long as parameters $p_{mi}$ have wrong values. In the process of learning, these parameters attain their correct values and, at the end of the DL-simulator processes, the association variables can be interpreted as probabilities.

These association variables are used to update parameter values, which is the second part of the DL process. In this case parameter update equations are simple,

$$p_{mi} = \sum_{n \in N} f(m|n) \, x_{ni} \, / \sum_{n' \in N} f(m|n')  \qquad (3)$$

These equations have very simple interpretations: they estimate parameters $p_{mi}$ of the m-th simulator as weighted averages of the bottom-up signals, $x_{ni}$. Note, the bottom-up signals "do not know" which situation they came from. The weights ($f/\Sigma f$) are normalized association variables, associating data n with simulator m. These equations are easy to understand: if the object i never appears in a situation m, at the end of the DL-simulator learning process, $f(m|n) = 0$, and $p_{mi} = 0$, as it should be, even if $x_{ni}$ are not 0 because object i appears in other situations. The role of the normalizing denominator ($\Sigma f$) is easy to understand; for example, if object i is actually present in situation m, then $x_{ni} = 1$ for each set of bottom up signal n, whenever situation m is observed. In this case, at the end of DL-simulator process, $f(m|n) = 1$ for all n, and eq.(3) yields $p_{mi} = 1$, as it should be.

For shortness, we did not discuss relations among objects. Spatial, temporal, or structural connections, such as "to the left," "on top," or "connected" can be easily added to the above DL formalism. Relations and corresponding markers (indicating which objects are related) are no different mathematically than objects, and can be considered as included in the above formulation. This mechanism is "flat" in the hierarchical structure of the brain, meaning that relations "reside" at the same level as entities they relate. Alternatively, some relations are realized in the brain hierarchically: relations could "reside" at a higher level, with markers being implemented similar to parametric models. Experimental data might help to find out, which relations are "flat" and which are "hierarchical." Other types of relations are principally hierarchical, e.g. objects-features, situations-objects, etc. We would also add that some relations are not "directly observable", as objects; say to differentiate between "glued to" or "stack to" might require knowledge of human actions or how the world is organized. Prerequisites to some of this knowledge might be inborn (Pinker 2008). We suggest that directly observable relations are learned as parts of a situation, similar to objects, and this learning is modeled by the DL formalism described above. Relations that require human cultural knowledge may be learned with the help of language, as discussed later, and inborn mechanisms should be further elucidated experimentally. This discussion implies several predictions that could be experimentally tested: existence of two types of relation mechanisms, flat and hierarchical; suggestions of which types of mechanisms are likely to be used in the brain for which types of relations; and suggestions of mechanisms conditioned by culture and language.

We repeat that the formulation here assumes that all the objects have already been recognized, but the above formulation can be applied without any change to real, continuously working brain with multiplicity of concurrently running simulators at many levels, feeding each other. Also modality of objects (various sensorial or motor mechanisms) requires no modifications (emotions can be included as well, some emotions are reducible to representations and learned to be a part of a situation similar to objects; other involve entirely different mechanisms discussed later). The bottom up signals do not have to be definitely recognized objects, these signals can be sent before objects are fully recognized, while object simulators are still running and object representations are vague; this would be represented by $x_{ni}$ values between 0 and 1. The bottom up signals do not have to correspond to complete objects, but could recreate patterns of activations in sensorimotor brain areas associated with perception of objects; similarly, top-down signals corresponding to situations, $p_m$, correspond to patterns of activations recreating experience of this situation. The presented formalization therefore is a general mechanism of simulators. A fundamental experimentally testable prediction of the developed theory is that top-down signals originate from vague representations, and the vagueness is determined by degrees of uncertainty of association between bottom-up signals and various higher-level representations.

## 3.2.    Example of symbol-situation learning using DL

In this example we set the total number of recognizable objects equal to 1000 ($D_o$=1000). The total number of objects perceived in a situation is set to 50 ($D_p$=50). The number of essential objects is set to 10 ($D_s$=10). The number of situations to learn (M-1) is set to 10. Note that the true identities of the objects are not important in this simulation so we simply use object indexes varying from 1 to 1000 (this index points to neural signals corresponding to a specific object-simulators). The situation names are also not important and we use situation indexes (this index points to neural signals corresponding to a specific situation-simulators). We would emphasize that the use of numbers for objects and situation, while may seem consistent with amodal symbols, in fact is nothing but notations. We repeat that the principled differences between PSS and amodal systems are mechanisms in the brain and their modeling, not mathematical notations. Among these mechanisms are simulators, mathematically described by DL. Let us repeat, amodal symbols are governed by classical logic, which is static and faces CC. DL is a process and overcomes CC. DL operates on PSS representations (models $p_m$), which are vague collections of objects (some of these objects could also be vague, not completely assembled yet representations). Another principled difference is interaction between perceptual-based bottom-up and top-down neural fields $X_n$ and $M_m$; indexes n and m are just mathematical shorthand for corresponding neural connections. In this paper we consider object perception and situation perception in different sections, but of course the real mind-brain operates continuously, "objects" in this section are neural signals sent to situation-recognition brain area (and corresponding simulators) by excited neuron fields corresponding to models of recognized-objects (partially, as described in section 2; and as discussed, these signals are being sent before objects are fully recognized, while object simulators are still running).

The data for this example are generated by first randomly selecting $D_s$=10 specific objects for each of the 10 groups of objects, allowing some overlap between the groups (in terms of specific objects). This selection is accomplished by setting the corresponding probabilities $p_{mi}$ = 1. Next we add 40 more randomly selected objects to each group (corresponding to $D_p$=50). We also generate 10 more random groups of 50 objects to model situations without specific objects (noise); this is of course equivalent to 1 group of 500 random objects. We generate N'=800 perceptions for each situation resulting in N=16,000 perceptions (data samples, n = 1… 16,000) each represented by 1,000-dimensional vector $\mathbf{X}_n$. These data are shown in Fig. 2 sorted by situations.

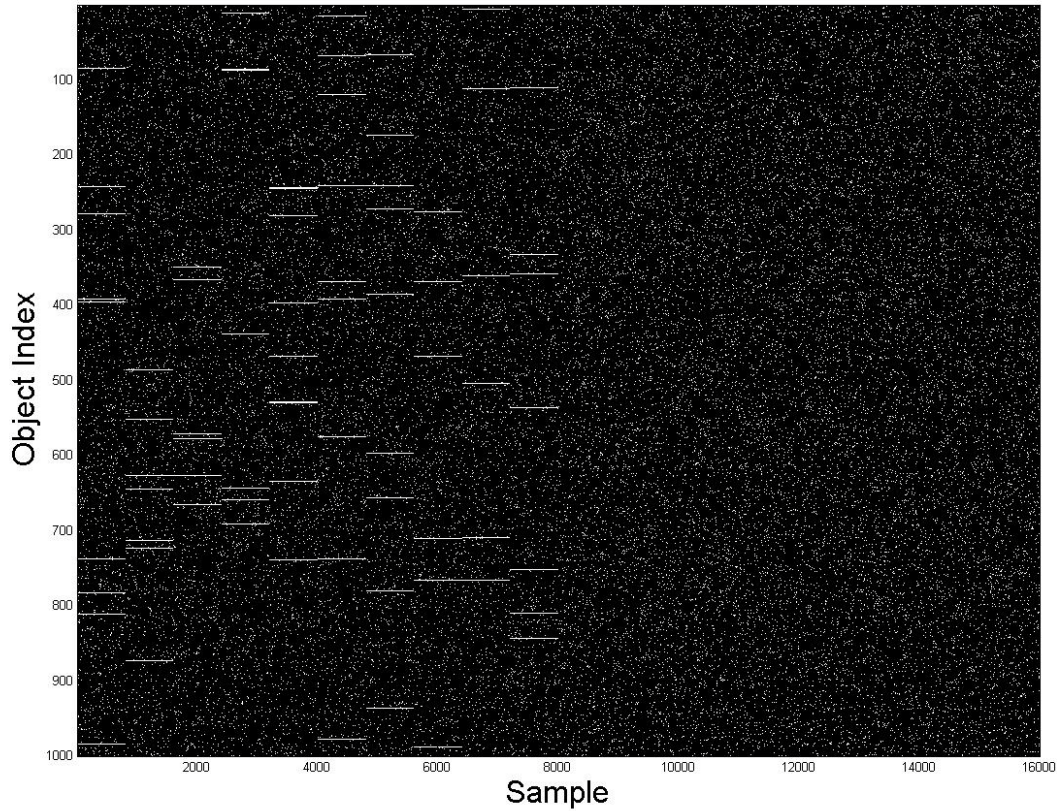

Fig. 2. Generated data; object index is along vertical axes and situation index is horizontal. The perceptions (data samples) are sorted by situation index (horizontal axis); this makes visible the horizontal lines for repeated objects.

Then the samples are randomly permuted, according to randomness of real life perceptual situations, in Fig. 3. The horizontal lines disappear; the identification of repeated objects becomes nontrivial. An attempt to learn groups-situations (the horizontal lines) by inspecting various horizontal sortings (until horizontal lines would become detectable) would require $M^N = 10^{16000}$ inspections, which is of course impossible. This CC is the reason why the problem of learning situations has been standing unsolved for decades. By overcoming CC, DL can solve this problem as illustrated below.

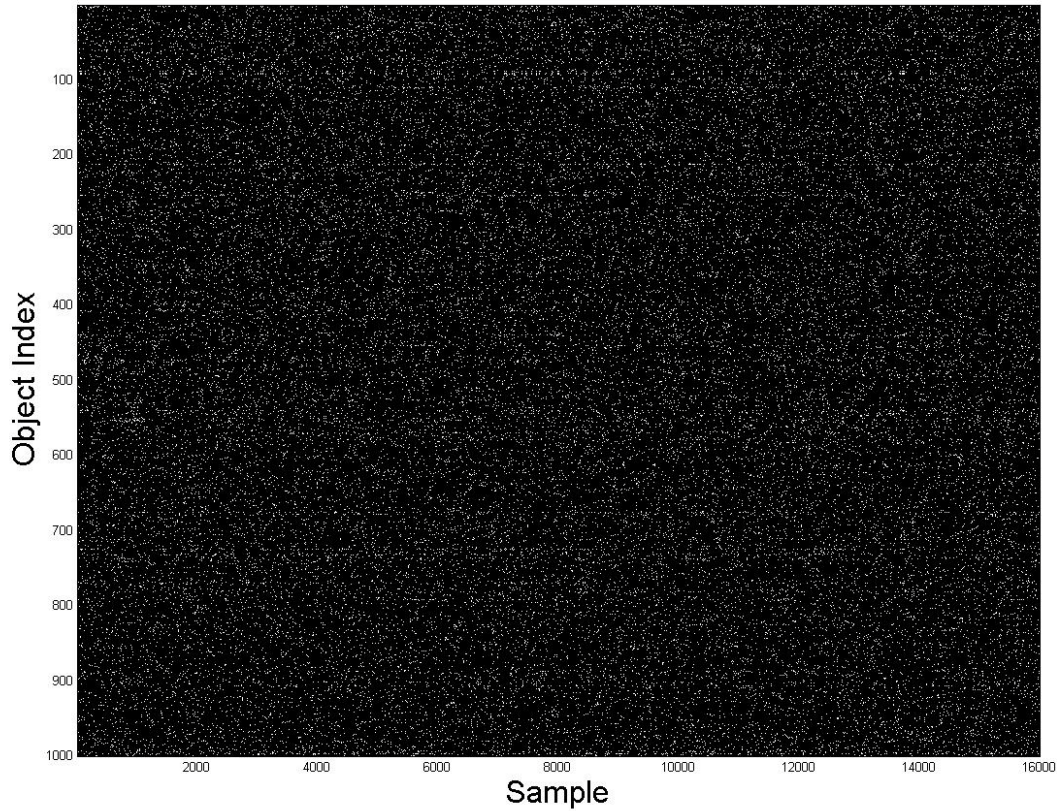

Fig. 3. Generated data, same as Fig. 2, randomly sorted by situation index (horizontal axis), as available to the DL algorithm for learning.

The DL algorithm is initiated similarly to section 2 by defining 20 situational models (an arbitrary selection, given actual 10 situations) and one random noise model to give a total of M=21 models (in section 2.4, Fig.1 models were automatically added by DL as required; here we have not done this (because it would be too cumbersome to present results). The models are initialized by assigning random probability values to the elements of the models. These are the initial vague perceptual models, which assign all objects to all situations.

Fig. 4 illustrates the initialization and the iterations of the DL algorithm (the first 3 steps of solving DL equations. Each subfigure displays the probability vector $\mathbf{p_m}$ for each of the 20 models. The vectors have 1000 elements corresponding to objects (vertical axes). The values of each vector element are shown in gray scale. The initial models assign nearly uniformly distributed probabilities to all objects. The horizontal axes are the model index changing from 1 to 20. The noise model is not shown. As the algorithm progresses, situation grouping improves, and only the elements corresponding to repeating objects in "real" situations keep their high values, the other elements take low values. By the third iteration the 10 situations are identified by their corresponding models. The other 10 models converge to more or less random low-probability vectors.

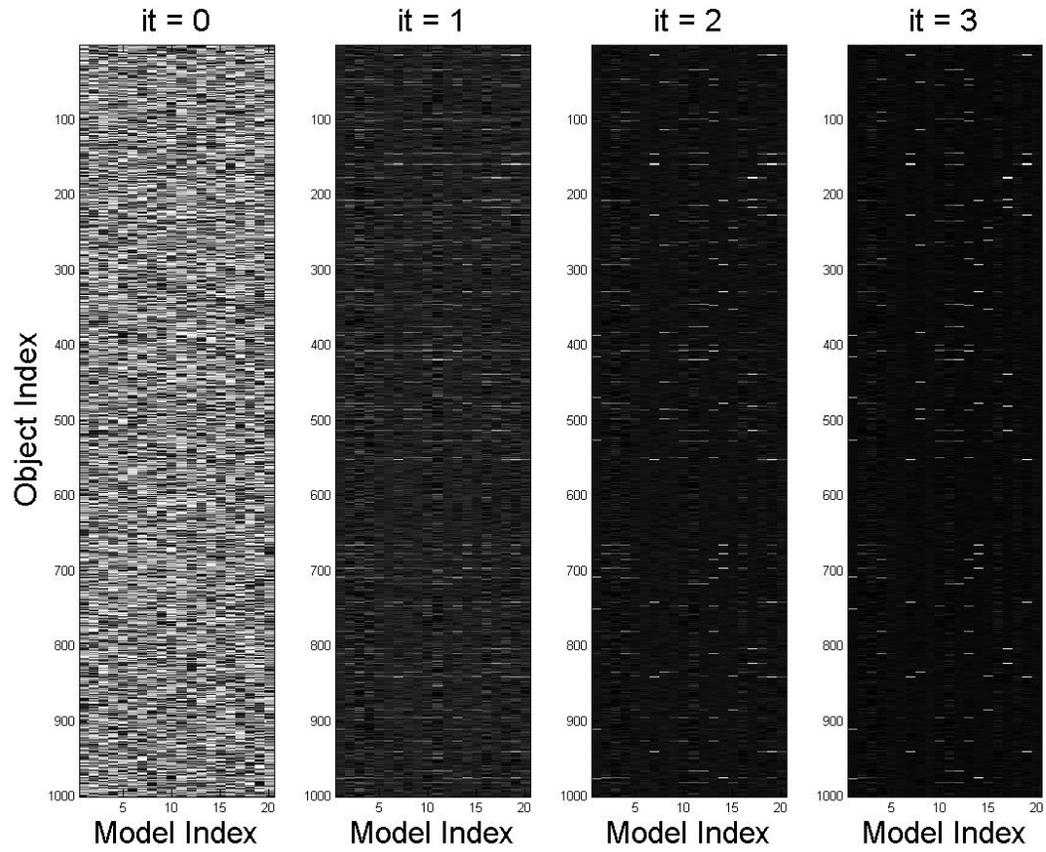

Fig. 4. DL situation learning. Situation-model parameters converge close to true values in 3 steps.

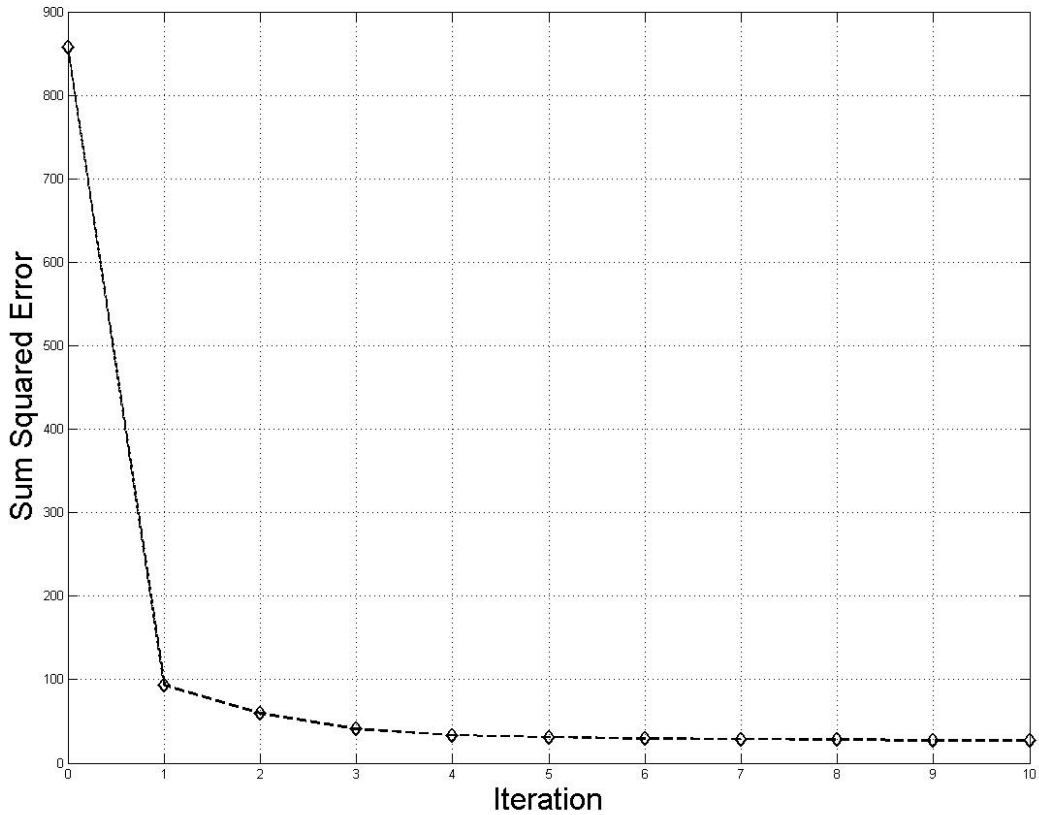

Fig. 5. Errors of DL learning are quickly reduced in 3-4 steps, iterations continue until average error reached low value of 0.05 (10 steps).

This fast and accurate convergence can be seen from Figs. 5 and 6. We measure the fitness of the models to the data by computing the sum-squared error, using the following equation.

$$ E = \sum_{m \in \{B\}} \sum_{i=1}^{Do} ( p_{mi} - p_{mi}^{True})^2 $$

In this equation the first summation is over the subset {B} containing top 10 models that provide the lowest error (and correspondingly, the best fit to the 10 true models). In the real brain, of course, the best models would be added as needed, and the random samples would accumulate in the noise model automatically; as mentioned, DL can model this process and the reason we did not model it, is that it would be too cumbersome to present results. Fig. 5 shows how the sum squared error changes over the iterations of the DL algorithm. It takes only a few iterations for the DL algorithm to converge. Each of the best models contains 10 large and 990 low probabilities. Iterations stop, when average error of probabilities reached a low value of 0.05 resulting in the final error E(10) =1000*(0.05^2 )*10 = 25.

Fig. 6 shows average associations, A(m,m') among true (m) and computed models (m'); this is an 11x11 matrix according to the true number of different models (it is computed using association variables between models and data, f(m|n))

$$A(m,m') = (1/N') \sum_{n=1}^{N} f(m|n)* f(m'|n), \ m' \in \{B\},$$

$$A(m,11) = (1/10*N') \sum_{m' \notin \{B\}} \sum_{n=1}^{N} f(m|n)* f(m'|n), \ m' \notin \{B\}$$

Here, f(m|n) for true 10 models m is either 1 (for N' data samples from this model) or 0 (for others), f(m'|n) are computed associations, in the second line all 10 computed noise models are averaged together, corresponding to one true (random) noise model. The correct associations on the main diagonal in Fig. 6 are 1 (except noise model, which is spread among 10 computed noise models, and therefore equals 0.1) and off-diagonal elements are near 0 (incorrect associations, corresponding to small errors shown in Fig 5.).

We would like to address briefly, why errors in Fig. 5 do not converge exactly to 0. The reason is numerical, p and x variables in eq. (1) cannot be allowed to take exactly 0 value, since expressions $0^0$ are numerically non-defined, therefore all values have been bounded from below by 0.05 (a somewhat arbitrary limit). Fig. 6 demonstrates that nevertheless, convergence to the global maximum was achieved (the exactly known solution in terms of learning the correct situations).

Again, as in section 2, learning of perceptual situation-symbols has been accomplished due to the DL process-simulator, which simulated internal model-representations of situations, **M**, to match patterns in bottom-up signals **X** (sets of lower-level perceptual object-symbols).

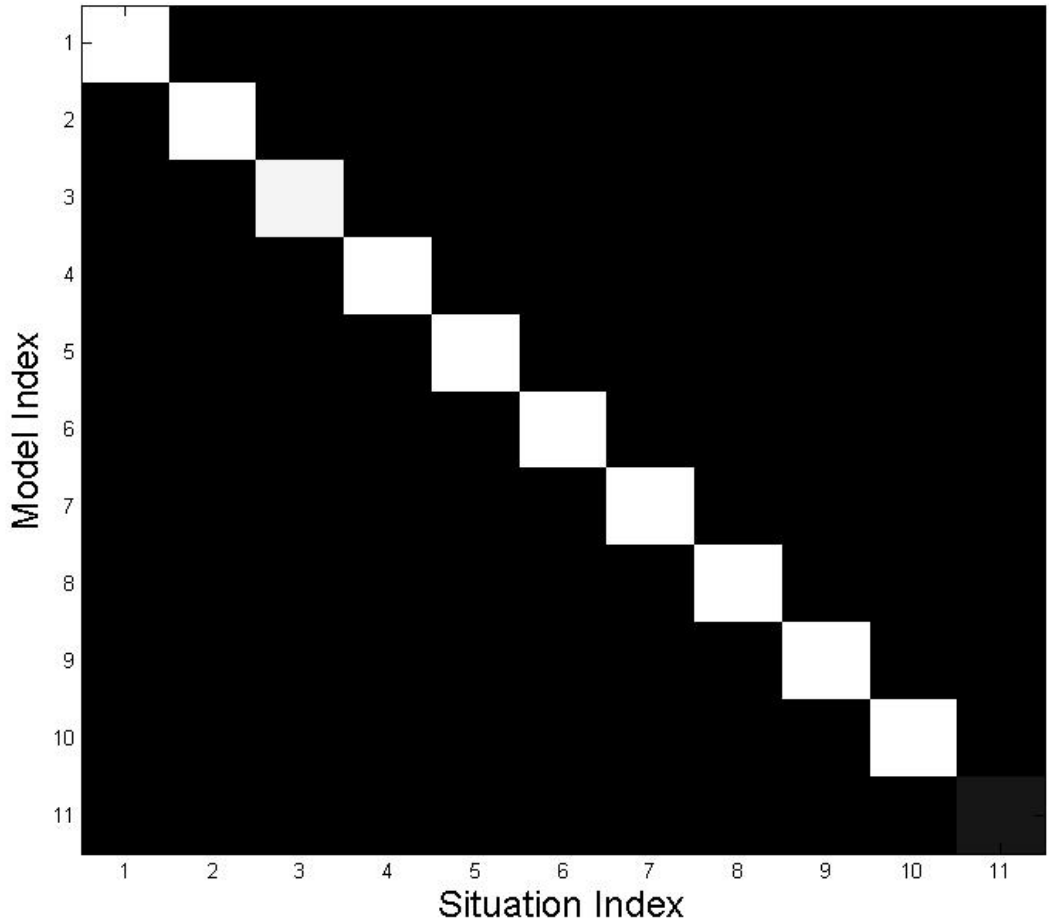

Fig. 6. Correct associations are near 1 (diagonal, except noise) and incorrect associations are near 0 (off-diagonal).

## 4. Simulators, concepts, grounding, binding, and DL

As described previously, PSS grounds perception, cognition, and high-level symbol operation in modal symbols, which are ultimately grounded in the corresponding brain systems. Previous section provides an initial development of formal mathematical description suitable for PSS: the DL process "from vague-to-crisp" models PSS simulators. We have considered just one subsystem of PSS, a mechanism of learning, formation, and recognition of situations from objects making up the situations. (More generally, the formalized mechanism of simulators includes recognition of situations by recreating patterns of activations in sensorimotor brain areas, from objects, relations, and actions making up the situations). The mind's representations of situations are symbol-concepts of a higher level of abstractness than symbol-objects making them up. The proposed mathematical formalism can be advanced straightforwardly to "higher" levels of more and more abstract concepts. We would add a word of caution: such application to more abstract ideas may require an additional grounding in language (Perlovsky 2007b; 2009a) as we briefly consider in the next section.

The proposed mathematical formalism can be similarly applied at a lower level of recognizing objects as constructed from their parts; mathematical techniques of sections 2 and 3 can be combined to implement this PSS object recognition idea as described in (Barsalou 1999). Specific models considered in section 2 are likely to be based on inborn mechanisms specific to certain aspects of various sensor and motor modalities; general models of section 3 can learn to represent and recognize objects as collections of multi-modal perceptual features and relations among them. In both cases principal mechanisms of object perception such as discussed in (Spelke 1990) can be modeled, either as properties of object models, or as relations between perceptual features. Since relations specific to object recognition, according to this reference are learned in infancy, the mechanism of section 3 seems appropriate (it models learning of relations, whereas models in section 2 do not readily contain mechanisms of learning of all their structural aspects and are more appropriate to modeling inborn mechanisms). Object representations, as described by Barsalou are not similar to photographs of specific objects, but similar to models in Fig.4 are more or less loose and distributed (among modalities) collections of features (determined in part by inborn properties of sensor and motor organs) and relations.

We note that the described theory, by modeling the simulators, also mathematically models productivity of the mind concept-simulator system. The simulated situations and other concepts are used not only in the process of matching bottom-up and top-down signals for learning and recognizing representations, but also in the motor actions, and in the processes of imagination and planning.

Modeling situations in PSS as a step toward general solution of the binding problem is discussed in (Edelman & Breen 1999). DL provides a general approach to the binding problem similar to the "corkboard" approach described in (Edelman & Intrator 2001). That publication also discusses the role of context similar to the DL scene modeling. Here we would emphasize two mechanisms of binding modeled in the developed theory. First, binding is accomplished hierarchically: e.g. object representations-simulators bind features into objects, similarly situation representations-simulators bind objects into situations, etc. Second, binding is accomplished by relations that are learned similarly to objects and "reside" at the same level in the hierarchy of the mind with the bound entities. These two types of binding mechanisms is another novel prediction of the DL theory that could be tested experimentally.

Below we discuss other relationships between the mathematical DL procedures of previous sections and the fundamental ideas of PSS. Section 2 concentrated on the principal mathematical difficulty experienced by all previous attempts to solve the problem of complex symbol formation from less complex symbols, the combinatorial complexity (CC). CC was resolved by using DL, a mathematical theory, in which learning begins with vague (non-specific) symbol-concepts, and in the process of learning symbol-concepts become concrete and specific. Learning could refer to a child's learning, which might take days or months or an everyday perception and cognition, taking approximately $1/6^{th}$ of a second (in the later case learning refers to the fact that every specific realization of a concept in the world is different in some respects from any previous occurrences, therefore learning-adaptation is always required; in terms of PSS, a simulator always have to re-assemble the concept). In the case of learning situations as compositions of objects, the initial vague state of each situation-symbol is a nearly random and vague collection of objects, while the final learned situation consists of a crisp collection of few objects specific to this situation. This specific of the DL process "from vague-to-crisp" is a prediction that can be experimentally tested, and we return to this later. In the learning process random irrelevant objects are "filtered out," their probability of belonging to a concept-situation is reduced to zero, while probabilities of relevant objects, making up a specific situation is increased to a value characteristic of this object being actually present in this situation.

Relation of this DL process to PSS is now considered. First we address concepts and their development in the brain. According to (Barsalou 2007),

> "The central innovation of PSS theory is its ability to implement concepts and their interpretative functions using image content as basic building blocks."

This aspect of PSS theory is implemented in DL in a most straightforward way. Concept-situations in DL are collections of objects (symbol-models at lower levels, which are neurally connected to neural fields of object-images). As objects are perceptual entities-symbols in the brain, concept-situations are collections of perceptual symbols. In this way situations are perceptual symbols of a higher order complexity than object-symbols, they are grounded in perceptual object-symbols (images), and in addition, their learning is grounded in perception of images of situations. A PSS mathematical formalization of abstract concepts (Barsalou 2003b), not grounded in direct perceptions, is considered in the next section. Here we just mention that the proposed model is applicable to higher levels, "beyond" object-situations; it is applicable to modeling interactions between bottom-up and top-down signals at every level.

Barsalou (2008) has described development of concepts in the brain as forming collections of correlated features. This is explicitly implemented in the DL process described in section 3. The developed mathematical representation corresponds to multimodal and distributed representation in the brain. It has been suggested that a mathematical set or collection is implemented in the brain by a population of conjunctive neurons (Simmons & Barsalou 2003).

DL learning and perception-cognition processes are mathematical models of PSS simulators. DL symbol-situations are not static collections of objects but dynamic processes. In the process of learning they "interpret individuals as tokens of the type" (Barsalou 2008). They model multi-modal distributed representations (including motor programs) as described in the reference.

The same DL mathematical procedure can apply to perception of a real situation in the world as well as an imagined situation in the mind. This is the essence of imagination. Models of situations (probabilities of various objects belonging to a situation, and objects attributes, such as their locations) can depend on time, in this way they are parts of simulators accomplishing cognition of situations evolving in time. If "situations" and "time" pertain to the mind's imaginations, the simulators implement imagination-thinking process, or planning.

Usually we perceive-understand a surrounding situation, while at the same time thinking and planning future actions and imagine consequences. This corresponds to running multiple simulators in parallel. Some simulators support perception-cognition of the surrounding situations as well as ongoing actions, they are mathematically modeled by DL processes that converged to matching internal representations (types) to specific subsets in external sensor signals (tokens). Other simulators simulate imagined situations and actions related to perceptions, cognitions, and actions, produce plans, etc.

Developed here DL modeling of PSS models mathematically what Barsalou (2003b) called dynamic interpretation of PSS (DIPSS). DIPSS is fundamental to modeling abstraction processes in PSS. Three central properties of these abstractions are type–token interpretation; structured representation; and dynamic realization. Traditional theories of representation based on logic model interpretation and structure well but are not sufficiently dynamical. Conversely, connectionist theories are dynamic but are inadequate at modeling structure. PSS addresses all three properties. Similarly, the DL mathematical process developed here addresses all three properties. In type–token relations "propositions are abstractions for properties, objects, events, relations and so forth. After a concept has been abstracted from experience, its summary representation supports the later interpretation of experience." Correspondingly in the developed mathematical approach, DL models a situation as a loose collection of objects and relations. Its summary representation (the initial model) is a vague and loose collection of property and relation simulators, which evolves-simulates representation of a concrete situation in the process of perception of this concrete situation according to DL. This DL process involves structure (initial vague models) and dynamics (the DL process).
.

### 5. DL and PSS: abstract concepts, language, the mind hierarchy, and emotions

Here we discuss DL as a general model of interacting bottom-up and top-down signals throughout the hierarchy-heterarchy of the mind-brain, including abstract concepts. The DL mathematical analysis suggests that modeling the process of learning abstract concepts has to go beyond PSS analysis in (Barsalou 1999). In particular, we discuss the role of language in learning abstract concepts (Perlovsky 2007b; 2009a) and connect it to the PSS mechanisms. The mind-brain is not a strict hierarchy, interactions across layers are present and this is some times addressed as heterarchy (Grossberg 1988). To simplify discussion we would use a term hierarchy.

Section 3 discussed assembling situation representations from object representations. This addresses interaction between top-down and bottom-up signals in two adjacent layers of the mind hierarchy. The mathematical description presented in section 3 addresses top-down and bottom-up signals and representations without explicit emphasis on their referring to objects or situations. Accordingly, we would emphasize here that the mathematical formulation in section 3 equally addresses interaction between any two adjacent layers in the entire hierarchy of the mind-brain, including high-level abstract concepts. DL overcomes the ubiquitous problem of CC, the presented DL mathematics is practically computable in a machine or mind. However, another fundamental aspect, grounding, remains questionable.

In the PSS formulation Barsalou assumed that higher level abstract concepts remain grounded since they are based on lower level grounded concepts, and down the hierarchy to perceptions directly grounded in sensory-motor signals. The DL modeling suggests that this aspect of the PSS theory has to be revisited for the following two reasons. First, each higher level is vaguer than a lower level. Several levels on top of each other would result in representations too vaguely related to sensory-motor signals to be grounded in them with any reliability. Second, the section 3 example is impressive in its numerical complexity, which significantly exceeds anything that has been computationally demonstrated previously. We would like to emphasize again that this is mostly due to overcoming difficulty of CC. Still statistically, learning of situations was based on these situations being present among the data with statistically sufficient information to distinguish them among each other and from noise. In real life however, human learn complex abstract concepts, such as "state," "law," "rationality," and many other abstract concepts, without statistically sufficient information been experienced (we return to this statement later and discuss its various aspects and a need for experimental verification). Now we would like to emphasize the role of language in learning abstract concepts.

Language is learned at all levels of the hierarchy of the mind-brain and cultural knowledge from surrounding language. Experience of talking with other people operates in significant way with "ready made" language concepts. This makes it possible for kids to talk about much of cultural contents by the age of five or seven. At this age kids can talk about many abstract ideas, which they cannot yet adequately use in real life. This suggests that language concepts and cognitive concepts are different. Language concepts are grounded in surrounding language at all hierarchical levels. But learning corresponding cognitive concepts grounded in life experience takes entire life. Learning language, like learning cognition can be modeled by DL. Linguists consider words to be learned by memorizing them (Pinker 1994). Learning meaningful phrases and syntax is similar to learning situations and relations among objects in section 3. Morphology is not unlike object composition.

As discussed above and in the given references a fundamental difference of language from cognition is grounding. Let us repeat, language is grounded in direct experience (of talking, reading) at all levels of the hierarchy, whereas cognition is grounded in direct perceptions only at the bottom of the hierarchy. It seems to be an inescapable conclusion that higher abstract levels of cognition are grounded in language. The detailed theory of interaction between cognition and language is considered in (Perlovsky 2002; 2004; 2006b,d; 2009a; Fontanari & Perlovsky 2007; 2008a;b; Fontanari, Tikhanoff, Cangelosi, Ilin, & Perlovsky 2009; Perlovsky & Fontanari 2006; Perlovsky & Ilin 2010). It leads to a number of verifiable experimental predictions, which we summarize in section 8. The main suggested mechanism of interaction between language and cognition is the dual model, which hypothesize that every mental model-representation has two neurally connected parts, language model and cognitive model. Language models are learned by simulator processes, similar to PSS simulators, however, "perception" in case of language refers to perception of language facts. Through neural connections between the two parts of each model, the early acquired abstract language models guide the development of abstract cognitive models in correspondence with cultural experience stored in language. The dual model leads to the dual hierarchy illustrated in Fig. 7.

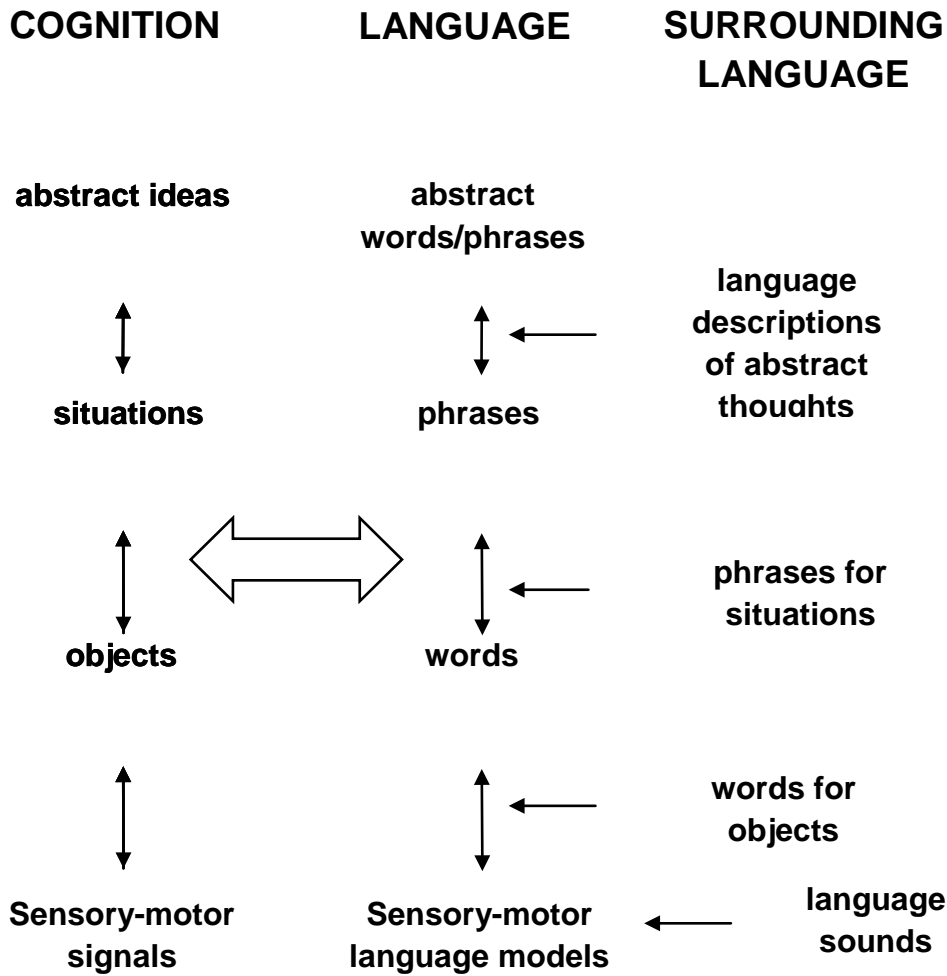

Fig. 7. The dual-model architecture, modeling interaction of language and cognition. Learning of cognition is grounded in experience and guided by language. Learning of language is grounded in the surrounding language at all hierarchical levels.

## 6. Perceptual vs. amodal symbols

Since any mathematical notation may look like an amodal symbol, in this section we discuss the roles of amodal vs. perceptual symbols in DL. This would require clarification of the word symbol. We touch on related philosophical and semiotic discussions and relate them to mathematics of DL and to PSS. For the sake of brevity within this paper we limit discussions to the general interest, emphasizing connections between DL, perceptual, and amodal symbols; extended discussions of symbols can be found in (Perlovsky 2006b;d). We also summarize here related discussions scattered throughout the paper.

"Symbol is the most misused word in our culture" (Deacon, 1998). Why the word "symbol" is used in such a different way: to denote trivial objects, like traffic signs or mathematical notations, and also to denote objects affecting entire cultures over millennia, like Magen David, Swastika, Cross, or Crescent? Let us compare in this regard opinions of two founders of contemporary semiotics, Charles Peirce (Peirce 1897; 1903) and Ferdinand De Saussure (1916). Peirce classified signs into symbols, indexes, and icons. Icons have meanings due to resemblance to the signified (objects, situations, etc.), indexes have meanings by direct connection to the signified, and symbols have meaning due to arbitrary conventional agreements. Saussure used different terminology, he emphasized that signs receive meanings due to arbitrary conventions, whereas symbol implies motivation. It was important for him that motivation contradicted arbitrariness. Peirce concentrated on the process of sign interpretation, which he conceived as a triadic relationship of sign, object, and interpretant. Interpretant is similar to what we call today a representation of the object in the mind. However, this emphasis on interpretation was lost in the following generation of scientists. This process of "interpretation" is close to the DL processes and PSS simulators. We therefore follow Saussurean designation of symbol as a motivated process. Motivationally loaded interpretation of symbols was also proposed by Jung (1921). He considered symbols as processes bringing unconscious contents to consciousness. Similar are roles of PSS simulators and DL processes. (Motivated in DL means in particular related to drives, emotions).

In the development of scientific understanding of symbols and semiotics, the two functions, understanding the language and understanding the world, have often been perceived as identical. This tendency was strengthened by considering logical rules to be the mechanism of both, language and cognition. According to Russell (1919), language is equivalent to axiomatic logic, "[a word-name] merely to indicate what we are speaking about; [it] is no part of the fact asserted… it is merely part of the symbolism by which we express our thought." Hilbert (1928) was sure that his logical theory also describes mechanisms of the mind, "The fundamental idea of my proof theory is none other than to describe the activity of our understanding, to make a protocol of the rules according to which our thinking actually proceeds." Similarly, logical positivism centered on "the elimination of metaphysics through the logical analysis of language" – according to Carnap (1959) logic was sufficient for the analysis of language. As discussed in section 2.2, this belief in logic is related to functioning of human mind, which is conscious about the final states of DL processes and PSS simulators; these final states are perceived by our minds as approximately logical amodal symbols. Therefore we identify amodal symbols with these final static logical states, signs.

DL and PSS explain how the mind constructs symbols, which have psychological values and are not reducible to arbitrary logical amodal signs, yet are intimately related to them. In section 3 we have considered objects as learned and fixed. This way of modeling objects indeed is amenable to interpreting them as amodal symbols-signs. Yet, we have to remember that these are but final states of previous simulator processes, perceptual symbols. Every perceptual symbol-simulator has a finite dynamic life, and then it becomes a static symbol-sign. It could be stored in memory, or participate in initiating new dynamical perceptual symbols-simulators. This infinite ongoing dynamics of the mind-brain ties together static signs and dynamic symbols. It grounds symbol processes in perceptual signals that originate them; in turn, when symbol-processes reach their finite static states-signs, these become perceptually grounded in symbols that created them. We become consciously aware of static sign-states, express them in language and operate with them logically. Then, outside of the mind-brain dynamics, they could be transformed into amodal logical signs, like marks on a paper. Dynamic processes – symbols-simulators are usually not available to consciousness. These PSS processes involving static and dynamic states are mathematically modeled by DL in section 3 and further discussed in section 4.

To summarize, in this paper we follow a tradition using a word sign for an arbitrary, amodal, static, unmotivated notation (unmotivated means unemotional, in particular). We use a word symbol for the PSS and DL processes-simulators, these are dynamic processes, connecting unconscious to conscious; they are motivationally loaded with emotions. As discussed in section 2, DL processes are motivated toward increasing knowledge, and they are loaded with knowledge-related emotions, even in absence of any other motivation and emotion. These knowledge-related emotions are called aesthetic emotions since Kant. They are foundations of higher cognitive abilities, including abilities for the beautiful, sublime, and they are related to musical emotions. More detailed discussions can be found in (Levine & Perlovsky 2008, Perlovsky 1996b, 1997, 2000, 2001, 2002a,c, 2004, 2006a,b,d, 2007a,b,c, 2009a,b, 2010a,b, Perlovsky, Bonniot-Cabanac, & Cabanac 2010, Perlovsky & Ilin 2010).

DL mathematical models (in section 3) use mathematical notations, which could be taken for amodal symbols. Such an interpretation would be erroneous. Meanings and interpretations of mathematical notations in a model depends not on the appearance, but on what is modeled. Let us repeat, any mathematical notation taken out of the modeling context, is a notation, a static sign. In DL model-processes these signs are used to designate neuronal signals, dynamic entities evolving 'from vague to crisp' and mathematically modeling processes of PSS simulators-symbols. Upon convergence of DL-PSS simulator processes, the results are approximately static entities, approximately logical, less grounded and more amodal.

DL models both, grounded, dynamic symbol-processes, overcoming combinatorial complexity and amodal static symbols, which are governed by classical logic and in the past have led to combinatorial complexity. DL operates on a non-logical type of PSS representations, which are vague combinations of lower-level representations. These lower-level representations are not necessarily complete images or events in their entirety, but could include bits and pieces of various sensor-motor modalities, memory states, as well as vague dynamic states from concurrently running simulators – DL processes of the on-going perception-cognition. (In section 3, for simplicity of presentation, we assumed that the lower-level object-simulators have already run their course and reached static states; however, the same mathematical formalism can model simulators running in parallel on multiple hierarchical levels.) The mind-brain is not a strict hierarchy, the same-level and higher-level representations could be involved along with lower levels. DL models processes-simulators, which operate on PSS representations. These representations are vague and incomplete, and DL processes are assembling and concretizing these representations. As described in several references by Barsalou, bits and pieces from which these representations are assembled could include mental imagery as well as other components, including multiple sensor, motor, and emotional modalities; these bits and pieces are mostly inaccessible to consciousness during the process dynamics.

DL also explains how logic and ability to operate amodal symbols originate in the mind from illogical operations of PSS: mental states approximating amodal symbols and classical logic appear as the end of the DL process-simulators. At this moment they become conscious static representations and loose that component of their emotional-motivational modality, which is associated with the need for knowledge (to qualify as amodal, these mental states should have no sources of modality, including emotional modality). The developed DL formalization of PSS, therefore suggests using a word signs for amodal mental states as well as for amodal static logical constructs outside of the mind, including mathematical notations; and to reserve symbols for perceptually grounded motivational cognitive processes in the mind-brain. Memory states, to the extent they are static entities, are signs in this terminology. Logical statements and mathematical signs are perceived and cognized due to PSS simulator symbol-processes and become signs after being understood. Perceptual symbols, through simulator processes, tie together static and dynamic states in the mind. Dynamic states are mostly outside of consciousness, while static states might be available to consciousness.

## 7. Experimental evidence

Bar et al. (2006) demonstrated in neuroimaging experiments that visual perception proceeds according to DL simulating crisp perceptions from initial vague representations (In this reference authors use terminology of "low spatial frequency" for what we call vague). Experimental procedures in this reference used functional Magnetic Resonance Imaging (fMRI) to obtain high-spatial resolution of processes in the brain, which they combined with magneto-encephalography (MEG), measurements of the magnetic field next to the head, which provided high temporal resolution of the brain activity. Combining these two techniques the experimenters were able to receive high resolution of cognitive processes in space and time. Bar et al. concentrated on three brain areas: early visual cortex, object recognition area (fusiform gyrus), and object-information semantic processing area (OFC). They demonstrated that OFC is activated 130 ms after the visual cortex, but 50 ms before object recognition area. Their conclusion has been that OFC represents the cortical source of top-down facilitation in visual object recognition. This top-down facilitation is unconscious. They demonstrated that the imagined image generated by top-down signals facilitated from OFC to cortex is vague (the authors in this publication refer to low spatial-frequency content images), confirming the essential mechanism of DL. Conscious perception of an object occurs when vague projections become crisp and match the crisp and clear image from the retina, and an object recognition area is activated.

The brain continuously extracts rudimentary information from early sensory data and simulates predictions, which facilitate perception and cognition in the relevant context by pre-sensitizing relevant representations. This includes predictions of complex information, such as situations and social interactions (Bar 2007; 2009). Predictions are initiated by gist information rapidly extracted from sensory data. At the "lower"-object level this gist information is a vague image of an object (low spatial frequency, Bar et al. 2006). At higher levels "the representation of gist information" is yet to be defined." The model developed here defines this higher-level gist information as vague collections of vague objects, with relevant objects for a specific situation having just slightly higher probabilities than irrelevant ones. The developed model is also consistent with the hypothesis in (Bar 2007) that perception and cognition at higher levels relies on mental simulations. Mathematical predictions in this paper suggest specific properties of these higher-level simulators (initial top-down representations are vague in terms of their associations with bottom-up signals), which could be verified experimentally.

## 8. Future research

Future research will address the DL mathematical description of PSS throughout the mind hierarchy; from features and objects "below situations" in the hierarchy to abstract models and simulators at higher levels "above situations." Modeling across the mind modalities will be addressed including diverse modalities, symbolic functions, conceptual combinations, predication. Modeling features and objects would have to account for suggestions that perception of features are partly inborn (Barsalou 1999); this development therefore might require new experimental data on which feature aspects are inborn (Edelman & Newell, 1998). The developed DL formalization of PSS corresponds to observations in (Wu & Barsalou 2009) and it will be used for generating more detailed experimentally verifiable predictions. A number of predictions have been made in this paper, including influence of perception on cognition (van Danzig et al, 2008).

The proposed theory provides solutions to classical problems of conceptual relations, binding, and recursion. Binding is a mechanism connecting events into meaningful "whole" (or larger-scale events). The DL model developed here specifies two types of binding mechanisms "flat" and "hierarchical," and suggests which mechanisms are likely to be used for various relations. Our model also suggests existence of binding mechanisms conditioned by culture and language. Recursion has been postulated to be a fundamental mechanism in cognition and language (Hauser, Chomsky, and Fitch, 2002), however, that reference has not proposed specific mechanisms how recursion creates representations, nor how it maps representations into the sensory-motor or conceptual-intentional interfaces. In our opinion this is an erroneous assumption, and the error is similar to mistaking logical signs for symbol-processes (recursion is an important logical operation). According to the developed theory recursion is not a fundamental mechanism, instead the hierarchy is proposed as a mechanism of recursion. Successive hierarchical levels accomplish recursive cognitive and linguistic functions. Again, these proposals can be experimentally tested.

Experimental research (Bar et al. 2006; Bar 2007) can address specific properties of higher-level simulators predicted here. Among these is a prediction that early predictive stages of situation simulations are vague. Whereas vague predictions of objects resemble low-spatial frequency of object imagery (Bar et al. 2006), "the representation of gist information on higher levels of analysis is yet to be defined" (Bar 2007). According to the developed model, vague predictions of situations should contain many less-relevant (and likely vague) objects with lower probabilities. Since the mathematical model proposed here is applicable to higher levels ("above" object-situations), this hypothesis should be relevant to the nature of information of higher-level gists: initial representation of abstract concepts are vague in terms of associations with constituent bottom-up signals (these associations are not exact, but vague; probabilistically, they are not close to zeroes and ones).

The present model can be expanded to address another topic discussed in (Bar 2007), "how the brain integrates and holds simultaneously information from multiple points in time." Two different mechanisms should be explored: first, explicit incorporation of time into models (so that model parameters-probabilities depend on time), and second, categorized temporal relations, such as "before," "after" can be included as any other relations-objects into models. A joint mathematical-experimental approach might be fruitful in this area.

Future research will address interaction between language and cognition. Language is acquired from surrounding language, rather than from direct experience in the world; language therefore is closer aligned with amodal symbols than with perceptual symbols. Kids at 5 years of age can talk about much of cultural content of the surrounding language, including highly abstract contents; yet, clearly kids do not have necessary experience to understand highly abstract concepts, as perceptual symbols, and to relate them to the world. According to the developed theory, higher abstract concepts could be stronger grounded in language than in perception; not only kids, but also adults may operate with abstract concepts as with amodal symbols, and therefore have limited understanding grounded in experience of how abstract concepts relate to the world. It follows that higher-level concepts may be less grounded in perception and experience than in language. The developed theory suggests several testable hypotheses: (i) the dual model, postulating separate cognitive and language mental representations, (ii) neural connections between cognitive and language mental representations, (iii) language mental representations guiding acquisition of cognitive representations in ontological development, (iv) abstract concepts being more grounded in language than in experience; and (v) this shift from grounding in perception and experience to grounding in language progresses up the hierarchy of abstractness; while grounding in perception and experience increases with age. These make a fruitful field for future experimental research.

The suggested model of connections between language and cognition bears on language evolution, and future research should address theoretical and experimental tests of this connection between evolution of languages, cognition, and cultures (Brighton, Smith, & Kirby 2005; Perlovsky & Fontanari 2006; Fontanari & Perlovsky 2007; 2008a;b; Fontanari, Tikhanoff, Cangelosi, Ilin, & Perlovsky 2009; Perlovsky 2009b).

The role of emotions in perception was addressed in (Barrett & Bar 2009). There are several mechanisms of emotions and future research should extend this paper formalism to more detailed modeling of emotions and their role in cognition. Also future research would explore roles of emotions (i) in language-cognition interaction (Perlovsky 2009b), (ii) in symbol grounding, and (iii) the role of aesthetic and musical emotions in cognition (Perlovsky 2000, 2000c, 2006c, 2008, 2010).

# Appendix 1.

Bottom-up signals $\{\mathbf{X}(n)\}$ in this simplified discussion, is a field of neuronal synapse activations in visual cortex. Here and below curve brackets $\{\ldots\}$ denote multiple signals, a field. Index n = 1,… N, enumerates neurons and $\mathbf{X}(n)$ are the activation levels. Sources of top-down signals are representations or concept-models $\{\mathbf{M}_m(n)\}$ indexed by m = 1,… M. Each model m $\mathbf{M}_m(n)$ projects a set of priming, top-down signals, representing the bottom-up signals $\mathbf{X}(n)$ expected from a particular object, m. Models depend on parameters $\{\mathbf{S}_m\}$, $\mathbf{M}_m(\mathbf{S}_m,n)$. Parameters characterize object position, angles, lightings, etc. (In case of learning situations considered in section 3, parameters characterize objects and relations making up a situation.) To summarize this highly simplified description of a visual system, n enumerates the visual cortex neurons, $\mathbf{X}(n)$ are the "bottom-up" activation levels of these neurons coming from the retina, and $\mathbf{M}_m(n)$ are the "top-down" activation levels (priming) of the visual cortex neurons. The learning-perception process "matches" these top-down and bottom-up activations by selecting "best" models and their parameters and the corresponding sets of bottom-up signals.

Let us concentrate on defining a mathematical measure of the "best" fit between bottom-up and top-down signals. It is constructed in such a way that any of object-models can be recognized.

Correspondingly, a similarity measure is designed so that it treats each object-model as a potential alternative for each subset of signals (Perlovsky 2000, 2006),

$$L(\{\mathbf{X}\},\{\mathbf{M}\}) = \prod_{n \in N} \sum_{m \in M} r(m) \, l(\mathbf{X}(n) \mid \mathbf{M}_m(n)); \qquad (A1)$$

Here, $l(\mathbf{X}(n)|\mathbf{M}_m(n))$ (or simply $l(n|m)$) is called a conditional similarity between one signal $\mathbf{X}(n)$ and one model $\mathbf{M}_m(n)$. Parameters $r(m)$ are proportional to the number of objects described by the model m. Expression (1) accounts for all combinations of signals and models in the following way. Sum $\sum$ ensures that any of the object-models can be considered (by the mind) as a source of signal $\mathbf{X}(n)$. Product $\prod$ ensures that all signals have a chance to be considered (even if one signal is not considered, the entire product is zero, and similarity L is 0; so for good similarity all signals have to be accounted for. This does not assume exorbitant amount of attention to each minute detail: among models there is a vague simple model for "everything else"). In a simple case, when all objects are perfectly recognized and separated from each other, there is just one object-model corresponding to each signal (other $l(n|m) = 0$). In this simple case expression (1) contains just 1 item, a product of all non-zero $l(n|m)$. In the general case, before objects are recognized, L contains a large number of combinations of models and signals; a product over N signals is taken of the sums over M models; this results in a total of $M^N$ items; this huge number is the cause for the combinatorial complexity discussed previously.

The DL learning process consists in estimating model parameters $\mathbf{S}_m$ and associating subsets of signals with concepts by maximizing the similarity (1). Although (1) contains combinatorially many items, DL maximizes it without combinatorial complexity (Perlovsky, 1996b; 1997; 2000). First, fuzzy association variables $f(m|n)$ are defined,

$$f(m|n) = r(m) \, l(n|m) \, / \sum_{m' \in M} r(m') \, l(n|m'). \qquad (A2)$$

These variables give a measure of correspondence between signal $\mathbf{X}(n)$ and model $\mathbf{M}_m$ relative to all other models, m'. They are defined similarly to the a posteriori Bayes probabilities, they range between 0 and 1, and as a result of learning they converge to the probabilities under certain conditions.

DL process is defined by the following set of differential equations,

$$df(m|n)/dt = f(m|n) \sum_{m' \in M} \{ [\delta_{mm'} - f(m'|n)] \, [\partial \ln l(n|m')/\partial \mathbf{M}_{m'}] \, (\partial \mathbf{M}_{m'}/\partial \mathbf{S}_{m'}) \, d\mathbf{S}_{m'}/dt,$$

$$d\mathbf{S}_m/dt = \sum_{n \in N} f(m|n)[\partial \ln l(n|m)/\partial \mathbf{M}_m]\partial \mathbf{M}_m/\partial \mathbf{S}_m, \quad \delta_{mm'} = 1 \text{ if } m=m', \text{ 0 otherwise.} \qquad (A3)$$